\documentclass[aps,reprint,superscriptaddress,
nofootinbib,showpacs,amsmath,amssymb]{revtex4-1}

\usepackage{latexsym,graphicx,slashed,bm,dcolumn}
\usepackage{hyperref}

\newcommand{\be}[1]{\begin{equation}\label{#1}}
\newcommand{\beq}{\begin{equation}}
\newcommand{\eeq}{\end{equation}}
\def\ee{\end{equation}}
\newcommand{\beqn}[1]{\begin{eqnarray}\label{#1}}
\newcommand{\eeqn}{\end{eqnarray}}

\newcommand{\dub}[2]{\left(\begin{array}{c}{#1}\\{#2}
\end{array}\right)}

\newcommand{\mat}[4]{\left(\begin{array}{cc}{#1}&{#2}\\{#3}&{#4}
\end{array}\right)}
\newcommand{\matr}[9]{\left(\begin{array}{ccc}{#1}&{#2}&{#3}\\
{#4}&{#5}&{#6}\\{#7}&{#8}&{#9}\end{array}\right)}

\renewcommand{\to}{\rightarrow}

\def\ov{\overline}

\def\lsim{\raise0.3ex\hbox{$\;<$\kern-0.75em\raise-1.1ex
\hbox{$\sim\;$}}}
\def\gsim{\raise0.3ex\hbox{$\;>$\kern-0.75em\raise-1.1ex
\hbox{$\sim\;$}}}
\def\cal{\mathcal}
\def\cF{{\cal F}}

\def\cL{{\cal L}}

\def\cN{{\cal N}}

\def\eps{\varepsilon}
\def\teps{\tilde\varepsilon}

\newcommand{\Dm}{\Delta m}

\def\rB{{\rm B}}
\def\rL{{\rm L}}

\begin{document}

\title{How light the lepton flavor changing gauge bosons can be?   }

\author{Benedetta Belfatto} 
\thanks{benedetta.belfatto@gssi.it} 
\affiliation{Gran Sasso Science Institute, Viale Francesco Crispi 7,  67100 'Aquila, Italy } 
\affiliation{INFN, Laboratori Nazionali del Gran Sasso, 67010 Assergi,  L'Aquila, Italy}

\author{Zurab~Berezhiani}
\email{zurab.berezhiani@lngs.infn.it}
\affiliation{Dipartimento di Fisica e Chimica, Universit\`a di L'Aquila, 67100 Coppito, L'Aquila, Italy} 
\affiliation{INFN, Laboratori Nazionali del Gran Sasso, 67010 Assergi,  L'Aquila, Italy}


\begin{abstract} 
Spontaneous breaking of inter-family (horizontal) gauge symmetries  
can  be at the origin of the mass hierarchy between the fermion families. 
The corresponding gauge bosons have flavor-nondiagonal couplings which 
generically induce the flavour changing phenomena,   
and this puts strong lower limits on the flavor symmetry breaking scales. 
However, in the special choices of chiral horizontal symmetries the flavor changing 
effects can be naturally suppressed. 
For the sake of demonstration, we consider the case of leptonic gauge  symmetry  $SU(3)_e$
acting between right-handed leptons and 
show that the respective  gauge bosons can have mass in the  TeV range, 
without contradicting the existing experimental limits. 
\end{abstract}

\maketitle


\noindent {\bf 1.} The replication of fermion families is one of the main puzzles of particle physics. 
%
Three fermion families are in identical representations of the Standard Model (SM) gauge symmetry 
$SU(3)\times SU(2)\times U(1)$. 
Its electroweak (EW) part $SU(2)\times U(1)$ is chiral with respect to fermion multiplets: 
the left-handed (LH) leptons and quarks, $\ell_{Li}=(\nu_i,e_i)_L$ and  $Q_{Li}=(u_i,d_i)_L$, 
transform  as weak isodoublets while the right-handed (RH) ones 
$e_{Ri},u_{Ri},d_{Ri}$ as isosinglets, $i=1,2,3$ being the family index. 
The chiral fermion content of SM has a remarkable feature that the fermion masses 
emerge only after spontaneous breaking of $SU(2)\times U(1)$ 
by the  vacuum expectation value (VEV) $\langle \phi^0\rangle = v_{\rm w} = 174$ GeV 
of the Higgs doublet $\phi$, via the Yukawa couplings 
 \be{Yukawas-SM}
Y_e^{ij} \, \phi \, \ov{\ell_{Li}} e_{Rj}  + Y_d^{ij} \, \phi  \, \ov{Q_{Li}} d_{Rj}  +
Y_u^{ij} \, \tilde\phi \, \ov{Q_{Li}} u_{Rj}  \, + \, {\rm h.c.} \, , 
\ee
where $Y_{e,u,d}$ are the Yukawa constant matrices, and $\tilde\phi=i\tau_2 \phi^\ast$.   
The fermion mass matrices $M_f = Y_f v_{\rm w}$, $f = e,u,d$, can be diagonalized via 
bi-unitary transformations $V_{Lf}^\dagger M_f V_{Rf} = M_f^{\rm diag}$. 
The masses of leptons $m_e,m_\mu,m_\tau$ and quarks $m_u,m_d,\dots$   
are the eigenvalues of these mass matrices. 
The ``right" matrices $V_{Rf}$ have no physical meaning in the SM context 
while the ``left" ones $V_{Lf}$ determine 
the mixing matrices in charged currents coupled to weak $W^\pm$ bosons,  
namely $V_{\rm CKM} = V_{Lu}^\dagger V^{\,}_{Ld}$ for quarks. 
However, no flavor mixing emerges in neutral currents coupled to $Z$ boson and Higgs boson.    
%
In this way, the SM exhibits a remarkable feature 
of natural suppression of flavor-changing neutral currents (FCNC) \cite{FCNC}:
all FCNC phenomena are suppressed at tree level and emerge exclusively from radiative corrections.  
At present, the majority of experimental data on flavor changing and CP violating 
processes are in good agreement with the 
SM predictions. There are few anomalies,  not definitely confirmed yet, 
which could point towards new physics beyond the Standard Model (BSM).     
 

In a sense, the SM is technically natural since it can tolerate any Yukawa matrices   $Y_f^{ij}$, 
but it tells nothing about  their  structures which remain arbitrary.       
So the origin of the fermion mass hierarchy and their weak mixing pattern 
remains a mystery.  
\medskip 

\noindent {\bf 2.}  The key for understanding the fermion mass and mixing pattern 
may lie in  symmetry principles. 
One can assign, e.g. different charges of abelian  flavor symmetry  
$U(1)$ to different fermion species  \cite{Froggatt},   
or one can introduce non-abelian horizontal gauge symmetries $G_H$ 
as e.g. $SU(3)_H$  \cite{Chkareuli,PLB,Khlopov} with  
the flavor gauge fields dynamically marking the family indices.  
Such a gauge theory of flavor can be considered as quantum flavordynamics,  
provided that it is built in a consistent way and sheds some more light 
on the origin of the fermion mass hierarchy. 

Namely, one can envisage that 
the form of the Yukawa matrices in (\ref{Yukawas-SM})  
is related to the VEV structures  of horizontal scalar fields (known also as flavons) 
which spontaneously break $G_H$, 
and the fermion mass hierarchy emerges from  the hierarchy between the 
scales of this breaking.  
 In Ref. \cite{PLB} this conjecture was coined as {\it hypothesis of horizontal hierarchies} (HHH).  
 It implies that the fermion masses cannot be induced without breaking $G_H$  
so that it cannot be a vector-like symmetry,  
but it should have a chiral character transforming the LH and RH particle species in different representations.  
In such a picture the fermion Yukawa couplings should emerge 
from the higher order ``projective" operators containing flavon scalars which transfer the VEV pattern  
of flavons to the structure of the Yukawa matrices $Y_f$. 
In the UV-complete pictures  such operators 
 can be induced  via renormalizable interactions after integrating out  some extra heavy fields, 
 scalars \cite{Chkareuli} or verctor-like fermions \cite{PLB,Khlopov}. 
In the context of supersymmetry,  such horizontal symmetries can lead to interesting relations 
between the  mass spectra of fermions and their superpartners 
and naturally  realize the minimal flavor violation scenarios \cite{MFV1,MFV2}.  

Discovery of the flavor gauge bosons of flavor and/or related FCNC effects  
would point towards new BSM physics of flavor. However, a  direct discovery 
at future accelerators can be realistic only if the scale of $G_H$ symmetry breaking 
is rather low, in the range of few TeV. 
Therefore, the following  questions arise: (i) for which choice of symmetry group $G_H$ 
 one can realize the HHH paradigm, relating the fermion mass hierarchy to its breaking pattern,  
and (ii) which is the minimal scale of $G_H$ symmetry 
 allowed  by present experimental limits,  and  namely, 
 can this scale be low enough to have $G_H$ flavor bosons 
within the potential experimental reach?

\medskip 

\noindent {\bf 3. }
In the limit of  vanishing Yukawa couplings,  $Y_f \to 0$, 
the SM acquires a maximal global chiral symmetry 
\be{max} 
U(3)_\ell \times U(3)_e \times U(3)_Q \times U(3)_u \times U(3)_d  \,  
\ee 
under which fermion species transform  
as triplets of independent $U(3)$ groups 
respectively as $\ell_{L}\sim 3_\ell$,   $ e_{R} \sim 3_e$, etc.  
The  Yukawa couplings (\ref{Yukawas-SM}) can be 
induced by the VEVs of flavons in mixed representations of these symmetry groups. 
 One can consider  the higher order operators  such as e.g. for leptons 
 \be{Ops-l}
\frac{X_e}{M} \phi \, \ov{\ell_{L}} e_{R} + {\rm h.c.} 
\ee
where $X_e \sim (3_\ell,\bar3_e)$ is a flavon in mixed representation 
of $U(3)_\ell \times U(3)_e$ 
which can be also viewed as composite tensor product $3_\ell \times \bar3_e$ 
of scalars in fundamental representations of $U(3)_\ell$ and $U(3)_e$.  

In the SM extensions the maximal flavor symmetry (\ref{max}) reduces to a smaller symmetry. 
E.g. in the context of $SU(5)$ grand unified theory (GUT) which unifies $\ell_L$ and $d^c_L$ fragments 
of each family in $\bar5$-plets and $e^c_L$, $u^c_L$ and $Q_L$  fragments in $10$-plets 
($\psi^c_L= C\ov{\psi_R}^T$, $C$ is a charge conjugation matrix), 
the maximal symmetry reduces to two factors $U(3)_\ell \times U(3)_e $: 
\be{su5}
\bar 5_L = (\ell, d^c)_L \sim (3_\ell, 1), ~  10_L = (e^c, u^c,Q)_L \sim (1, \bar3_e)  
\ee  
In the context of $SO(10)$ GUT all fermions of one family 
including the RH neutrino $N_R$
 reside in the spinor multiplet $16_L = (\bar5 + 10 + 1)_L$. 
Hence, there can be only one chiral symmetry $U(3)$ between three families of 16-plets, 
with all LH fermions $\ell_L,Q_L$ transforming 
as triplets and the RH ones $N_R,e_R,u_R,d_R$ as anti-triplets, in the spirit of chiral 
horizontal $SU(3)_H$ of Refs. \cite{Chkareuli,PLB,Khlopov}.  
For predictive models based on $SO(10) \times SU(3)_H$ see  e.g. in Refs. \cite{SO10}. 

It is tempting to consider some part of the maximal flavor symmetry (\ref{max}),  
or its GUT-restricted versions, 
as a gauge symmetry $G_H$. 
%
%
Gauging of chiral $U(1)$ factors  is  difficult since they are anomalous 
with respect to the SM.\footnote{ However,  there are models 
in which string-inspired 
anomalous  gauge symmetry $U(1)_A$ is used as a flavor symmetry  \cite{U1A}. } 
Therefore, we consider a situation in which only some of non-abelian  $SU(3)$ parts in (\ref{max}) 
are gauged.
%
In particular, in this paper we concentrate on the lepton sector and 
discuss a simple model with a gauge symmetry $G_H=SU(3)_e$  
transforming the RH leptons  as a triplet  $e_{R\alpha} = (e_1,e_2,e_3)_R$,  
while the LH leptons $\ell_{Li} = \ell_{1,2,3}$ have no symmetry and $i=1,2,3$ 
is just a family number.\footnote{Alternatively, one could say that 
also $SU(3)_\ell$ is a gauge symmetry but broken at some higher scales.    
More complete model with $SU(3)_\ell \times SU(3)_e$ symmetry
will be discussed elsewhere \cite{BB-new}.
}
We show that the lepton mass hierarchy $m_\tau \gg m_\mu \gg m_e$ 
can be directly related to the hierarchy of $U(3)_e$ symmetry breaking scales. 
As for the lepton flavor violating (LFV) phenomena induced by $SU(3)_e$ gauge bosons,  
we show that they are strongly suppressed since the intermediate $SU(2)_e$ subgroup 
acts as an approximate custodial symmetry.\footnote{Also the vector-like 
$SU(2)$ acting on both LH and RH fermion species  has custodial properties  \cite{Dvali}. 
However, it allows degenerate mass spectrum 
which makes problematic the naturalness of inter-family mass hierarchy.}   
The respective scale is allowed to be as low as 2 TeV, 
without contradicting the present experimental limits on the LFV processes.\footnote{
For comparison, the naive lower limit on the scale of flavor changing bosons is over 100 TeV 
\cite{Cahn}. In the models \cite{Chkareuli,PLB,Khlopov} this scale 
was assumed to be close to the GUT scale, and in any case larger than a PeV,  
for avoiding the excessive FCNC. 
For an exception, see Ref. \cite{Low}.}

\medskip 

\noindent {\bf 4.} 
The LH and RH lepton fields of our model are in the following representations: 
\be{reps} 
\ell_{Li} = \dub{\nu_i}{e_i}_L \sim (2,-1,1),   \quad \quad e_{R\alpha} \sim (1,-2,3_e)  
\ee 
where we explicitly indicate the multiplet content with respect to the EW $SU(2) \times U(1)$ 
and horizontal $SU(3)_e$.
This set of fermions is not anomaly free. The ways of the anomaly cancellations 
will be discussed in next sections.
  
We assume that there is only one Higgs doublet $\phi$  with the standard Higgs potential 
$V(\phi) = \lambda (\vert \phi\vert^2 - v_{\rm w}^2)^2$. 
However,  the Yukawa couplings of $\phi$ with the fermions $\ell_{Li}$ and $e_{R\alpha}$ 
are forbidden by $SU(3)_e$ symmetry.
So, for generating the lepton masses this symmetry should be broken. 

\begin{figure}[t]
\includegraphics[width=0.3\textwidth]{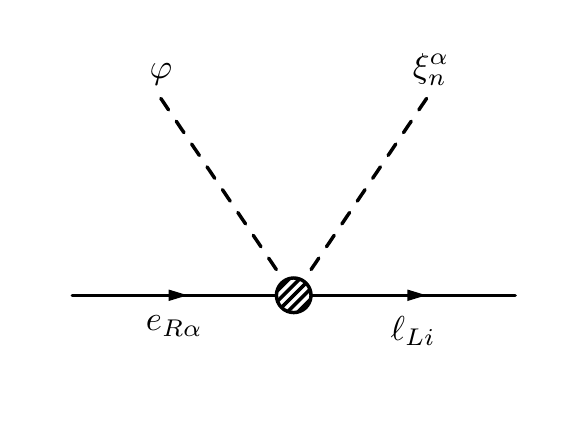} \\
\vspace{-6mm}
\includegraphics[width=0.45\textwidth]{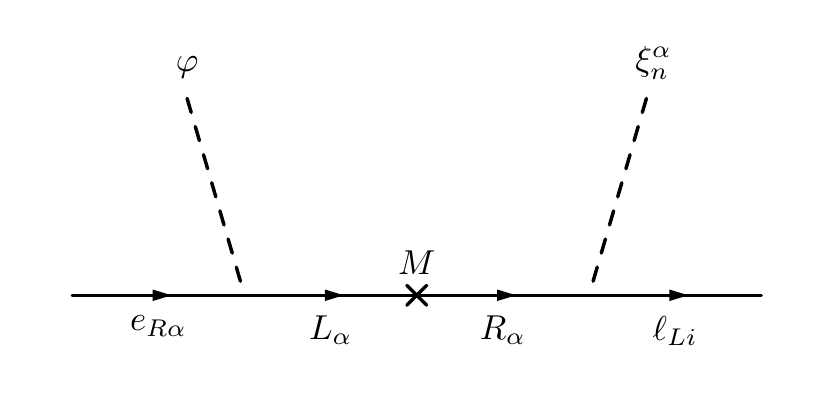}
\caption{ Upper diagram represents operator (\ref{op-leptons}) and lower diagram shows how it can be 
obtained via seesaw exchange of heavy vector-like fermions (\ref{Vectorlike}). 
}
\label{fig:seesaw}
\end{figure}

For breaking $SU(3)_e$ we introduce three flavon fields 
$\xi_n^{\alpha}$, $n=1,2,3$,  each transforming as $SU(3)_e$ (anti)triplet. 
The charged lepton masses then can emerge from the gauge invariant dimension 5  
operators 
\be{op-leptons} 
\sum_n \frac{g_{in} \xi_{n}^{\alpha} }{M }\,   \phi \, \overline{ \ell_{Li}  }  e_{R\alpha} \, + \, {\rm h.c.}
\ee 
where $g_{in}$ are order one constants (see upper diagram of Fig. \ref{fig:seesaw}).  
For having an UV-complete theory, one 
can consider these operators as induced form the renormalizable terms
via seesaw-like mechanism \cite{PLB}.   
E.g. one can integrate out from the following Yukawa Lagrangian   
\be{Yuk} 
h \phi \ov{L_\alpha} e_{R\alpha} + M \ov{R_\alpha} L_\alpha + 
\sum_n g_{in} \xi_n^\alpha \ov{\ell_{Li}} R_\alpha   
~+~ {\rm h.c.} \, , 
\ee
the extra heavy vector-like lepton doublets   
 \be{Vectorlike} 
L_\alpha, R_\alpha  = \dub{N_\alpha}{E_\alpha}_{\! L, R}    
 \sim (2,-1,3_e)   
\ee 
with a large Dirac mass  $M$ (see Fig. \ref{fig:seesaw}, lower diagram). 

Operator (\ref{op-leptons}) has a global symmetry $U(3)_e= SU(3)_e \times U(1)_e$, 
where the abelian part $U(1)_e$ is related to the phase change 
of fermions $e_{R\alpha}$ and flavons $\xi_n^{\alpha}$.  
%
%
%
In order to generate non-zero masses of all three leptons $e,\mu,\tau$,  this global symmetry must  
be fully broken. This means  that all three flavons  $\xi_n$ should have the non-zero VEVs  
with disoriented directions.  In other words, the VEVs $\langle \xi_n^\alpha \rangle$ 
should form a rank-3 matrix, which is generically non-diagonal.    
Then, without loss of generality, one can choose superpositions of these fields 
$\xi_n \to U_{nm} \xi_m$ 
so that their VEVs are orthogonal and hence the matrix $\langle \xi_n^\alpha \rangle$ becomes diagonal, 
$\langle \xi_{n}^{\alpha} \rangle = v_n \delta_{n}^{\alpha}$, 
or explicitly 
\be{VEV}
\langle \xi_1 \rangle = \left(\begin{array}{c}
v_1 \\ 0 \\ 0 
\end{array}\right), ~~
\langle \xi_2 \rangle = \left(\begin{array}{c}
0 \\ v_2 \\ 0
\end{array}\right)  , ~~
\langle \xi_3 \rangle = \left(\begin{array}{c}
0 \\ 0 \\ v_3
\end{array}\right) , 
\ee 
ordered as $v_3 > v_2 > v_1$ reflecting the steps of the global symmetry  
breaking $U(3)_e \to U(2)_e \to U(1)_e \to {\rm nothing}$. 
By substituting these VEVs 
in operator (\ref{op-leptons}), it reduces to the SM Yukawa couplings 
\be{SM-Yuk} 
Y_e^{i\alpha} \, \phi \, \ov{\ell_{Li}} e_{R\alpha} + {\rm h.c.},  \quad 
Y_e^{i\alpha} = \sum_n  \frac{g_{in} \langle \xi_n^\alpha \rangle}{M}  = g_{i\alpha} \frac{v_\alpha}{M} \, . 
\ee
Without loss of generality, $\ell_{Li}$ states can be turned to the basis in which matrix $g_{in}$ 
has a triangular form and the diagonal elements $g_{33},g_{22},g_{11}$ are real. Then
\beqn{Ye} 
 Y_e &= & \frac{1}{M} 
 \matr{g_{11} v_1 }{0 }{0 }
{g_{21} v_1 }{g_{22} v_2 }{0}
{g_{31} v_1 }{g_{32} v_2 }{g_{33} v_3 }  \nonumber \\
& = & \frac{v_3}{M} \matr{g_{11} \teps\eps }{0 }{0 }
{g_{21}\teps\eps }{g_{22} \eps }{0}
{g_{31} \teps\eps }{g_{32}\eps}{g_{33}} 
\eeqn
where we denote $v_2/v_3=\eps$ and $v_1/v_2 = \teps$.  
%
The Yukawa matrix $Y_e$ (and the mass matrix $M_e = Y_e v_{\rm w}$) 
can be diagonalized  via bi-unitary transformation 
\be{mixing}
Y_e  \to 
V_L^\dagger Y_e V^{\,}_R ={\rm diag} \big(y_e,y_\mu,y_\tau \big). 
\ee
Hence, 
the lepton mass hierarchy $m_\tau : m_\mu : m_e$  corresponds to 
the hierarchy between the scales $v_3 : v_2 : v_1$. 
Namely, neglecting the small $\sim \eps^2$ corrections, we have 
\be{emutau}
m_\tau = \frac{g_{33} v_3}{M} v_{\rm w}, ~~ m_\mu = \frac{g_{22} v_2}{M} v_{\rm w}, ~~
m_e = \frac{g_{11}  v_1}{M} v_{\rm w}\, . 
\ee

Let us discuss  whether such a hierarchy of the VEVs can be natural. 
Since three flavons have identical quantum numbers, 
their scalar potential has a generic form  
\beqn{V-flavon} 
{\cal V}(\xi) & = & 
 \lambda_n \Big(\big\vert \xi_n \big\vert^2  -\frac{\mu_n^2}{2\lambda_n}\Big)^2  
 + \lambda_{klnm}   \xi_k^\dagger \xi_l  \xi_n^\dagger \xi_m 
\nonumber \\
&&  
+ \, (\mu \xi_1\xi_2\xi_3 + {\rm h.c.} ) 
\eeqn
We assume that all constants $\lambda$ are 
say in the range $\lambda \sim 0.1\div1$,  and the mass terms  $\mu_n^2$, 
positive or negative, are of the same order, say within $1\div 10$ TeV. 
The last (trilinear) coupling $\mu \epsilon_{\alpha\beta\gamma} \xi_1^\alpha\xi^\beta_2\xi^\gamma_3$ 
has a dimensional constant $\mu$ which is however allowed (by 't Hooft's naturalness principle) 
to be arbitrarily small  
since in the limit $\mu\to 0$ the Lagrangian acquires global $U(1)_e$ symmetry 
respected also by the Yukawa terms  (\ref{op-leptons}). 
In fact, this latter coupling softly breaks $U(1)_e$ and thus reduces 
the global  symmetry $U(3)_e$ to $SU(3)_e$. 

For full breaking of gauge $SU(3)_e$ symmetry,  just two flavons  with non-aligned VEVs 
are sufficient. 
An order of magnitude hierarchy  between the scales $v_2$ and $v_3$,   
 $v_2/v_3 \sim m_\mu/m_\tau$,   can emerge due to some moderate conspiracy 
of parameters admitting a natural  ``spread", say within an order of magnitude,  between 
the mass terms and coupling constants of $\xi_2$ and $\xi_3$  in (\ref{V-flavon}). 
But large hierarchy  $v_1/v_3 \sim m_e/m_\tau$  at first sight requires a strong fine tuning. 
However, in fact small $v_1$ can be obtained naturally  considering the following situation.  
In the limit $\mu\to 0$ the VEV matrix $\langle \xi_n^\alpha \rangle$  has rank 2, 
so that only  two flavons $\xi_2$ and $\xi_3$  get the VEVs  
$v_{n} = \mu_{n}/\sqrt{2\lambda_n}$, $n=2,3$, oriented as in (\ref{VEV}), 
because of their negative mass$^2$ terms in (\ref{V-flavon}).   
 As for third flavon $\xi_1$, it has a positive mass$^2$ term,  i.e. $\mu_1^2< 0$, 
 and  in the limit $\mu=0$ it remains VEVless.  
However,  for $\mu\neq 0 $, the last term in (\ref{V-flavon})  
explicitly breaks global $U(1)_e$ symmetry and induces non-zero 
VEV $\langle \xi_1 \rangle $,  $v_1 = \mu v_2 v_3/\mu_1^2$. 
Thus, taking e.g.  $\mu_1\sim v_3$,  we have $ v_1/v_2 \sim \mu/v_3$,  and  
$\mu < v_2$ would suffice for having $v_1$ in the needed range.

The unitary matrix $V_R$ in (\ref{mixing}) 
connecting the initial flavor basis of the RH leptons  to the mass basis,   
\begin{equation}\label{VR}
\left(\begin{array}{c}
e_1\\e_2\\e_3
\end{array}\right)_{\!\!\!R} \!\! =V_R \! \left(\begin{array}{c}
e\\ \mu \\ \tau
\end{array}\right)_{\!\!\!R} \! =\left(\begin{array}{ccc}
V_{1e} & V_{1\mu} & V_{1\tau} \\ V_{2e} &V_{2\mu}& V_{2\tau}\\ V_{3e}& V_{3\mu}& V_{3\tau}
\end{array}\right) \!\!
\left(\begin{array}{c}
e\\ \mu \\ \tau
\end{array}\right)_{\!\!\!R}
\end{equation}
has no physical meaning for the EW interactions, but it  is meaningful for 
the LFV interactions mediated by the gauge bosons of $SU(3)_e$. 
Since the mixing angles in $V_R$ are small, 
we have (modulo $\eps^2$ corrections) $V_{1e}, V_{2\mu}, V_{3\tau} =1$ while    
for non-diagonal elements we get 
 \be{R-mix} 
 V_{3\mu} = - \frac{g_{32}}{g_{33}} \eps, \quad V_{2e} = - \frac{g_{21}}{g_{22}} \teps, 
 \quad V_{3e} = - \frac{g_{31}}{g_{33}} \teps \eps \, . 
 \ee
The other elements in matrix $V_R$ can be obtained from its unitarity. 

As for the mixing matrix $V_L$  of the LH charged leptons in (\ref{mixing}), 
it  is very close to unit matrix, with non-diagonal elements $\sim \eps^2$.  
Therefore, the neutrino mixing angles are determined  by the form of the 
neutrino mass matrix. 
The neutrino masses are induced by higher order operator \cite{Weinberg}
\be{Yukawa-nu} 
\frac{Y_\nu^{ij}  }{\cal M} \, \phi \phi  \, \ell^T_{Li} C \ell_{Lj}   \, + \, {\rm h.c.}  \, .
\ee
where ${\cal M}$ is a new scale which in the context of seesaw mechanism 
can be related to the Majorana  masses of RH neutrinos. 
In our scenario the states $\ell_{Li}$ are not distinguished by any symmetry 
and the matrix $Y_\nu^{ij}$ is a generic  non-diagonal matrix, 
supposedly  with all elements of the same order.  
 Thus, the unitary matrix $V_\nu$ which diagonalizes it, 
 $V_\nu^T Y_\nu V_\nu = Y_\nu^{\rm diag}$, contains large rotations 
 and the neutrino mixing angles are expected to be large. 

 \medskip 

\noindent{\bf 5.} 
Gauge bosons $\cF^\mu_a$ of $SU(3)_e$,  associated to the Gell-Mann matrices 
$\lambda_a$, $a=1,2,...8$,  interact as $g \cF^\mu_a J_{a\mu}$ 
with the respective currents 
$J_{a\mu} = \frac12 \ov{{\mathbf e}_R} \lambda_a \gamma_\mu {\mathbf e}_R$,
where $g$ is the $SU(3)_e$ gauge coupling and 
$\mathbf{e}_R = (e_1,e_2,e_3)_R^T$ denotes the triplet  of the RH leptons.  
Clearly, these currents are generically FCNC;  e.g. those related to non-diagonal 
generators $\lambda_{1,2}$ transform $e_1$ into $e_2$, etc. 
Nevertheless, as we shall see below, the processes mediated by flavor bosons exhibit no 
LFV in the basis of eigenstates $e_{R1}$, $e_{R2}$, $e_{R3}$  of  flavor  diagonal generators 
$\lambda_3$ and $\lambda_8$. 

At low energies the flavor bosons induce four-fermion (current $\times$ current)  interactions: 
\be{CC}
\mathcal{L}_{\rm eff} = - g^2\,  J_{a}^{\mu}\, (2M^2)^{-1}_{ab} \, J_{b\mu}  
\ee
where 
$M^2_{ab}$ is the (symmetric) mass matrix of gauge bosons $\cF^\mu_a$.  
%
In the flavon VEV basis (\ref{VEV})  this matrix is essentially diagonal 
apart of  a non-diagonal $2\times 2$ block related to ${\cal F}_3$ - ${\cal F}_8$ mixing. 
Namely, for the masses of gauge bosons $\cF^\mu_{4,5}$, $\cF^\mu_{6,7}$ and $\cF^\mu_{1,2}$ 
we have  respectively 
\beqn{67}
&& M_{4,5}^2 = \frac{g^2}{2}(v_3^2 + v_1^2), 
\quad M_{6,7}^2 = \frac{g^2}{2}(v_3^2 + v_2^2), \nonumber \\
&& M_{1,2}^2 = \frac{g^2}{2}(v_2^2 + v_1^2), 
\eeqn
while for the mass matrix of ${\cal F}^\mu_3$ - ${\cal F}^\mu_8$ system we get 
\be{38} 
M_{38}^2  =  \frac{g^2}{2} 
\mat{ v_2^2 + v_1^2} {\frac{1}{\sqrt3}(v_1^2 - v_2^2)}
 {\frac{1}{\sqrt3}(v_1^2 - v_2^2)} {\frac13(4 v_3^2 + v_1^2 + v_2^2) } \, .
 \ee
Obviously, the factor $g^2$ in operators (\ref{CC}) cancels 
and their strength is determined solely by 
$SU(3)_e$ symmetry breaking scales $v_2$ and $v_3$. 
In the following we neglect a small contribution $v_1^2/v_2^2 = \teps^2 \ll 1$ 
in the gauge boson mass terms and in respective effective operators. 
Then 
\be{inv-38} 
g^2(2M_{38}^2)^{-1} = \frac{1}{v_2^2} \mat{1}{0}{0}{0} + \frac{1}{4v_3^2} \mat{1}{\sqrt3}{\sqrt3}{3}
\ee 
Hence, from operators (\ref{CC})
one can single out the one cut off by the scale $v_2$: 
%
\be{G2}
{\cal L}_2 = - \frac{1}{v_2^2}   \sum_{a=1}^3 (J^\mu_a)^2  = 
- \frac{1}{4v_2^2} \sum_{a=1}^3 \, (\ov{\mathbf{e}_R} \, \lambda_a \gamma^\mu \mathbf{e}_R)^2    
\ee 
which involves only $e_{R1}$ and $e_{R2}$ states.  Using Fierz identities for $\lambda_{1,2,3}$   
which in fact are the Pauli matrices, this operator can be rewritten as  
\be{G2-cust}
{\cal L}_2  = - \frac{1}{v_2^2} \,  (J^\mu_0)^2 = 
- \frac{1}{4v_2^2} \, (\ov{ \mathbf{e}_R} \, \lambda_0  \gamma^\mu \mathbf{e}_R)^2  ,
\ee
where we denote $\lambda_0={\rm diag}(1,1,0)$.  
The remaining operators in (\ref{CC}) are related to the scale $v_3$: 
\beqn{G3} 
{\cal L}_3  = && 
- \frac{1}{v_3^2} \left[ (\tilde{J}_3^\mu)^2  + (J^\mu_4)^2   + (J^\mu_5)^2  \right] 
\nonumber \\
&&  \quad  - \frac{1}{v_3^2+ v_2^2} \left[ ({J}_6^\mu)^2  + (J^\mu_7)^2 \right] , 
\eeqn
where the current $\tilde{J}^\mu_3 = \frac12 J^\mu_3 + \frac{\sqrt3}{2} J^\mu_8$ has a form
$\frac12 \ov{ \mathbf{e}_R} \tilde\lambda_3 \gamma^\mu  \mathbf{e}_R$  
with $\tilde\lambda_3 = {\rm diag}(1,0,-1)$.  Hence, $\tilde\lambda_3$, $\lambda_4$ and $\lambda_5$ 
form a $SU(2)$ subalgebra of $SU(3)_e$, and using the Fierz identities for these matrices 
one can rewrite (\ref{G3}) as 
\beqn{G3-cust} 
&& {\cal L}_3  = - \frac{1}{v_3^2} \left[ 
(\tilde{J}_0^\mu)^2  + (J_6^\mu + i J_7^\mu)(J_{6\mu} - i J_{7\mu})   \right]  = \nonumber \\
&& - \frac{\eps^2}{4v_2^2} 
\left[(\ov{ \mathbf{e}_R} \, \tilde\lambda_0  \gamma^\mu \mathbf{e}_R)^2  + 
4 (\ov{e_{R2}}\gamma_\mu e_{R2})(\ov{e_{R3}}\gamma^\mu e_{R3}) \right] 
\eeqn 
where $\tilde\lambda_0={\rm diag}(1,0,1)$. 
So, operators $\cL_2$ and $\cL_3$ do not induce any LFV transition  between 
$e_{R1},e_{R2},e_{R3}$ states.  


In basis of mass eigenstates $(e,\mu,\tau)$
all currents involved in operators (\ref{G2-cust}) and (\ref{G3-cust}),  
including those related to $\lambda_0$ and $\tilde\lambda_0$, 
should be rotated with the matrix $V_R$ (\ref{VR}),  
or in explicit form  
\beqn{current} 
{J}_{a\mu} =  
\big(\ov{e},\ov{\mu},\ov{\tau} \big)_R 
\gamma_{\mu}\, \frac{V_R^{\dag} \lambda_a V_R}{2}  \left(\begin{array}{c}
e\\ \mu \\ \tau
\end{array}\right)_{\!\!R} 
\eeqn
In particular,  in $(e,\mu,\tau)$ basis $\lambda_0$ in operator (\ref{G2-cust}) is deformed to 
$\lambda_V = V_R^\dagger \lambda^{\,}_0 V^{\,}_R$, or explicitly   
\beqn{lambda} 
 \lambda_V  &= &
\matr{1-|V_{3e}|^2 }{ -V_{3e}^\ast V_{3\mu} } {-V_{3e}^\ast V_{3\tau} }
{ -V_{3e}V_{3\mu}^\ast } {1-|V_{3\mu}|^2 } {-V_{3\mu}^\ast V_{3\tau} }
{-V_{3e} V_{3\tau}^\ast  }{ - V_{3\mu} V_{3\tau}^\ast } { |V_{1\tau}|^2 + |V_{2\tau}|^2 } 
\nonumber \\ 
&& = \matr{1}{\sim \teps\eps^2 }{\sim \teps\eps }
{\sim \teps\eps^2 } {1} {\sim \eps }
{\sim \teps\eps } {\sim \eps }  {\sim \eps^2} \, .
\eeqn 
and $\tilde\lambda_V = V_R^\dagger \tilde\lambda^{\,}_0 V^{\,}_R$:
\beqn{tilde-lambda} 
 \tilde\lambda_V  &= &
\matr{1-|V_{2e}|^2 }{ -V_{2e}^\ast V_{2\mu} } {-V_{2e}^\ast V_{2\tau} }
{ -V_{2e}V_{2\mu}^\ast } {|V_{1\mu}|^2 + |V_{3\mu}|^2 } {-V_{2\mu}^\ast V_{2\tau} }
{-V_{2e} V_{2\tau}^\ast  }{ - V_{2\mu} V_{2\tau}^\ast } { 1- |V_{2\tau}|^2 } 
\nonumber \\ 
&& = \matr{1}{\sim \teps }{\sim \teps\eps }
{\sim \teps  } {\sim \eps^2 } {\sim \eps }
{\sim \teps\eps } {\sim \eps }  {1} \, .
\eeqn 
where we indicate the order of magnitude for the LFV entries  in terms of the small parameters 
$\eps$ and $\teps$.  

\medskip 

\noindent {\bf 6.}  
Let us question now how small the scale $v_2$ can be 
without contradicting to existing experimental limits.
The leading terms in (\ref{G2-cust}) give rise to flavor conserving operators 
\be{compo} 
- \frac{1}{4v_2^2} \big(\ov{e_R} \gamma^\nu e_R\big)^2 - \frac{1}{2v_2^2}  
\big(\ov{e_R} \gamma^\nu e_R\big) \big(\ov{\mu_R} \gamma_\nu \mu_R\big)
\ee
constrained by the compositeness limits  
 $\Lambda_{RR}^-(eeee) > 10.2$~TeV  
and $\Lambda_{RR}^-(ee\mu\mu) > 9.1$~TeV 
as reported respectively   in Refs. \cite{Bourilkov:2001pe} and \cite{Schael:2006wu}.  
Translating the formal definitions of compositeness scales to the scale $v_2$,  we get 
\beqn{comp-eeee} 
&&v_2 = (8\pi)^{-1/2} \Lambda_{RR}^-(eeee) > 2.0~{\rm TeV}  \nonumber \\
&& v_2 = (8\pi)^{-1/2} \Lambda_{RR}^-(ee\mu\mu) > 1.8~{\rm TeV} 
\eeqn 
Hence, these limits allow the scale  of $SU(2)_e$ symmetry to be as 
small as $v_2=2$~TeV or so. 
As for the flavor-changing phenomena induced by operators (\ref{G2-cust}) and ({\ref{G3-cust}),  
they are suppressed by small parameters $\eps$ and 
$\teps$ and agree with severe experimental limits on the LFV 
even for such a low scale of flavor symmetry.

\begin{table*}
\begin{tabular}{ l @{\makebox[5mm]} l @{\makebox[5mm]} l @{\makebox[5mm]} l }
\rule[-4mm]{0mm}{10mm}
LFV mode  & Experimental Br.   &  
\quad \quad  Predicted Br. &
   \\
\hline
\rule[-4mm]{0mm}{10mm}
$\mu\rightarrow ee\bar e$&
$ < 1.0\times 10^{-12}$&
$\frac18 \left(\frac{v_{\rm w}}{v_2}\right)^4\left\vert V_{3e}^*V_{3\mu} +
\eps^2 V_{2e}^\ast V_{2\mu} \right\vert^2$ &
$ \leq 1.1 \times 10^{-13}\left(\big\vert \frac{g_{31}^\ast g_{32}}{g_{33}^2} \big\vert +  
 \big\vert \frac{g^\ast_{21}}{g_{22}} \big\vert \right)^2  \teps_{20}^2 \eps_{20}^4  \left(\frac{2\,\text{TeV}}{v_2}\right)^4 $ \\
  \hline
 \rule[-4mm]{0mm}{10mm}
 $\tau \to \mu e\bar e$ &
$ < 1.8\times 10^{-8}$ &  $\frac14 \big(\frac{v_{\rm w}}{v_2}\big)^4
\left\vert  V_{3\mu}^\ast V_{3\tau} \right\vert ^2 \, {\rm Br}(\tau\to \mu\nu_\tau\bar\nu_\mu)$ &
$= 6.2 \times 10^{-9 }   \big\vert \frac{g_{32}}{g_{33}} \big\vert^2  
\eps_{20}^2  \left(\frac{2\,\text{TeV}}{v_2}\right)^4 $ \\
  \hline
 \rule[-4mm]{0mm}{10mm}
 $\tau\rightarrow\mu\mu\bar \mu$  &
$ < 2.1\times 10^{-8}$  &
$  \frac18 \big(\frac{v_{\rm w}}{v_2}\big)^4  
\big\vert V_{3\mu}^\ast V_{3\tau} \big\vert ^2 \, {\rm Br}(\tau\to \mu\nu_\tau\bar\nu_\mu)$ &
$= 3.1 \times 10^{-9 }  \big\vert \frac{g_{32}}{g_{33}} \big\vert^2 
\eps_{20}^2  \left(\frac{2\,\text{TeV}}{v_2}\right)^4 $ \\
\hline
\rule[-4mm]{0mm}{10mm}
$\mu\rightarrow e \gamma$&
$<4.2\times 10^{-13}$ &
$\frac{3\alpha}{2\pi} \big(\frac{v_{\rm w}}{v_2}\big)^4   \vert V_{3e}^\ast V_{3\mu} \vert^2 $ &
$= 3.1\times 10^{-15} \big\vert \frac{g_{31}^\ast g_{32}}{g_{33}^2} \big\vert^2 
 \teps_{20}^2\eps_{20}^4 \left(\frac{2\,\text{TeV}}{v_2}\right)^4 $\\
 \hline
\rule[-4mm]{0mm}{10mm}
$\tau\rightarrow \mu \gamma$&
$<4.4\times 10^{-8}$&
$\frac{3\alpha}{2\pi} \big(\frac{v_{\rm w}}{v_2}\big)^4   
\left\vert V_{3\mu}^\ast V_{3\tau} \right\vert ^2 \, {\rm Br}(\tau\to \mu\nu_\tau\bar\nu_\mu)$ &
$ = 8.7 \times 10^{-11} \big\vert \frac{g_{32}}{g_{33}} \big\vert^2  \eps_{20}^2 \left(\frac{2\,\text{TeV}}{v_2}\right)^4$
\end{tabular}
\caption{\label{Table} Experimental limits on the branching fractions for the LFV decays \cite{PDG} 
vs. those predicted in our model. 
 For the LFV decays of $\tau$ lepton  
the branching ratio ${\rm Br}(\tau\to \mu\nu_\tau\bar\nu_\mu) = 0.174$ is taken into account. 
In last column we used Eqs. (\ref{R-mix})  for the elements of mixing matrix $V_R$,  
set the scale $v_2=2$~TeV from the compositeness limits and parameters 
$\eps_{20} = 20\eps$,  $\teps_{20} = 20\teps$. 
}
\end{table*}

E.g. both $\cL_2$ and $\cL_3$ terms contribute  the following operator 
which induces the LFV decay $\mu \to ee\bar e$: 
\beqn{mu3e}
&&  \frac{4G_{\mu eee}}{\sqrt{2}}(\overline{e_R}\gamma_{\nu}\mu_R)\, (\overline{e_R}\gamma^{\nu}e_R) , 
\nonumber \\
&& \frac{4G_{\mu eee}}{\sqrt{2}}= \frac{1}{2v_2^2} V^\ast_{3e}V_{3\mu} 
+ \frac{1}{2v_3^2}  V_{2e}^\ast V_{2\mu}  \sim \frac{\eps^2 \teps}{2v_2^2}  \, ,  
\eeqn 
Its amplitude can be normalized  to the amplitude of the muon standard decay 
 due to the weak interactions 
\be{SM-mu} 
-\frac{4G_F}{\sqrt2} (\ov{e_L} \gamma_\rho \nu_{e}) \, (\ov{\nu_\mu} \gamma^\rho \mu_L), \quad 
\ee
where $4G_F/\sqrt2 = 1/v_{\rm w}^2$, $v_{\rm w}=174$~GeV. 
Hence, we get 
\be{Br-3e}
\frac{\Gamma({\mu \to ee\bar e})}{\Gamma(\mu \to e \bar\nu_e \nu_\mu)} 
 = \frac12 \left\vert\frac{G_{\mu eee}}{G_F}\right\vert^2 \sim 
\frac{\teps^2 \eps^4}{8} \left(\frac{v_{\rm w}}{v_2}\right)^4 \, .
\ee
Hence, for $v_2 > 2$~TeV  and $\eps,\teps \leq 1/20$ or so, this branching ratio is 
compatible with the existing experimental limit ${\rm Br}({\mu \to 3e})_{\rm exp} < 10^{-12}$ \cite{PDG}.  

For $\tau$ lepton decay modes as $\tau \to \mu e \bar{e}$ 
and $\tau \to 3\mu$  leading contributions arise from operator (\ref{G2-cust}). 
 From (\ref{lambda}) we get the relevant constants as 
$4G_{\tau \mu ee}\sqrt2 =4G_{\tau \mu\mu\mu}/\sqrt2  = V^\ast_{3\mu}V_{3\tau}/2v_2^{2}$. 
Hence,  the widths of these decays are suppressed by a factor 
 $\sim \eps^2/v_2^4$,  
 and are compatible with  the experimental limits \cite{PDG}. 
The summary of predicted branching ratios of relevant LFV processes 
compared with experimental limits is given in Table \ref{Table}, with the parameters 
$\eps$, $\teps$  normalized to a benchmark value $1/20$. 

Yet another LFV phenomena to be considered is muonium-antimuonium conversion
$M(\bar\mu e) \to \ov{M}(\mu \bar e) $ \cite{Feinberg}. 
The relevant operator emerging from (\ref{G3-cust}) reads 
\beqn{Muonium} 
 -\frac{4G_{M\ov{M} } }{\sqrt2} \big (\ov{\mu}_R \gamma^\nu e_R\big)^2, \quad 
G_{M\ov M} = \frac{\eps^2 (V_{2e} V_{2\mu}^\ast)^2 }{8\sqrt2 v_2^2} 
\eeqn 
Thus the amplitude of $M-\ov{M}$ transition is doubly suppressed, by a factor 
$\sim \teps^2 \eps^2 < 10^{-5}$ or so, 
and is much below the experimental limit $\vert G_{M\ov M}/G_F \vert < 3\times 10^{-3}$ \cite{Willmann}. 

One loop contribution of flavor bosons 
 to the electron magnetic moment has no suppression by mixing angles in $V_R$  
 (the electric dipole gets no contribution at one loop). 
By computing parameter $a_e= \frac12(g_e-2)$ with formulas in Ref. \cite{ae}, we get: 
\be{ae} 
a_e = - \frac{m_e^2}{8\pi^2 v_2^2} = 
-8.3 \times 10^{-16}  \left( \frac{2~{\rm TeV}}{v_2} \right)^2 
\ee
which is about  3 orders of magnitude smaller than the present difference between the 
experimental \cite{PDG} and theoretical \cite{Parker} 
determinations of the electron anomalous magnetic moment, 
$a_e^{\rm exp} - a_e^{\rm SM} = (-7.0 \pm 3.5) \times 10^{-13}$. 
Similarly, the contribution  for muon anomalous magnetic moment 
obtained by substituting $m_e \to m_\mu$ in (\ref{ae}), 
$a_\mu = -3.5 \times 10^{-11} (2~{\rm TeV}/v_2)^2$,  
is two orders of magnitude below the existing discrepancy 
$a_\mu^{\rm exp} - a_\mu^{\rm SM} = (2.7 \pm 0.8) \times 10^{-9}$ \cite{PDG}. 
So, these contributions are irrelevant for both electron and muon. 

Let us remark that potentially also flavons  can mediate the LFV processes. 
From the effective operators (\ref{op-leptons}), after substituting the VEV $\langle \phi \rangle = v_{\rm w}$ 
 we obtain  for the lepton Yukawa couplings  with the flavon fields $\xi_n$: 
\be{Yuk-flavon}
h_{in} \, \xi_n^\alpha \ov{\ell_{Li}} e_{R\alpha}, \quad h_{in} = \frac{g_{in} v_{\rm w}}{M} 
\ee
which are generically flavor-changing. 
For example, in the basis (\ref{Ye}) the Higgs mode of the flavon $\xi_2$ which is presumably the lightest, 
with the mass $\mu_2 \sim v_2$,  induces the following 
effective operator: 
\be{op-flavon} 
-\frac {h_{32} h_{22}}{\mu_2^2} \, (\ov{\tau} \mu ) \, (\ov{\mu} \mu ), \quad 
\frac {h_{32} h_{22}}{\mu_2^2}   \simeq  
\frac {m_{\mu}^2}{v_2^4} 
\ee
 where we have taken into account the relations (\ref{emutau}). Thus, for $v_2 > 2$ TeV, 
 the width of $\tau \to 3 \mu$ decay induced by this operator is  more than 12 orders of magnitude below
 the experimental limit.  The width of $\mu \to 3e$ decay induced by analogous operator 
 mediated by flavon $\xi_1$ is also suppressed by many orders of magnitude. 

\medskip

\noindent{\bf 7.} 
Let us remark that for promoting the chiral non-abelian factors in (\ref{max})  as $SU(3)_e$, etc. a
s gauge symmetries, one has to take care of anomaly cancellations. 
For more generality, let us consider 
\be{max-1} 
SU(3)_\ell \times SU(3)_e \times SU(3)_Q \times SU(3)_u \times SU(3)_d  \,  
\ee 
as maximal gauge symmetry  
under which the different fermion species form the triplets of independent $SU(3)$ 
horizontal groups 
respectively as   
\be{fermions}
\ell_{L}\sim 3_\ell,   ~ e_{R} \sim 3_e , ~  Q_{L}\sim 3_Q , ~ u_{R} \sim 3_u, ~ d_{R} \sim 3_d   \, . 
\end{equation} 
Then each of these gauge factors would have triangle $SU(3)^3$ anomalies.   
For their cancellation, for each triplet in (\ref{fermions}) one must additionally introduce
{\it ad hoc} fermions of the opposite 
chiralities which are the SM singlets and are  triplets under respective 
horizontal symmetry. 

Most interesting possibility is to introduce mirror sector \cite{Mirror} 
as a mirror copy of the SM gauge symmetry  $SU(3)\times SU(2) \times U(1)$, so that 
for every (LH or RH)  fermion  species of the SM, there exists its mirror twin with the opposite 
chirality in the identical representation of the mirror SM' gauge group $SU(3)'\times SU(2)' \times U(1)'$ 
(for a review see e.g.  Ref. \cite{Alice,Alice2}), and assume that horizontal symmetries (\ref{max-1}) 
are common symmetries between ordinary and mirror particles as it was suggested in Ref. \cite{MFV1}.  
In other words, for ordinary quark and lepton species  in (\ref{fermions}), 
their mirror twins (`primed'  quarks and leptons)  should be respectively in representations 
 \be{fermions-pr}
\ell'_{R}\sim 3_\ell,   ~ e'_{L} \sim 3_e , ~  Q'_{R}\sim 3_Q , ~ u'_{L} \sim 3_u, ~ d'_{L} \sim 3_d   \, .  
\end{equation} 
In this picture, the parity can be understood as a discrete symmetry of 
exchange between ordinary and mirror species, $\ell_L \leftrightarrow \ell'_R$ etc. 
with respective exchange of ordinary and mirror gauge bosons and Higgses $\phi$ and $\phi'$.
As for the horizontal gauge factors (\ref{max-1}), they in fact all become vector-like, 
with their triangle anomalies  reciprocally cancelled between the ordinary (\ref{fermions}) and 
mirror (\ref{fermions-pr}) particle species of the opposite chiralities.

In particular, in our  model ``reduced" to leptons in which only $SU(3)_e$ is considered as 
a gauge symmetry, triangle  $SU(3)_e^3$ anomaly is cancelled between the 
ordinary RH leptons $e_{R\alpha}\sim (1,-2,3_e)$  and their LH mirror partners 
$e'_{L\alpha}\sim (1,-2',3_e)$, where $-2'$ denotes $U(1)'$  hypercharge of mirror leptons. 

Operators (\ref{op-leptons}) must be complemented by similar couplings of flavons 
$\xi_n$ with mirror leptons.  Hence, we have  
\be{op-leptons-pr} 
\sum_n \frac{g_{in} \xi_{n}^{\alpha} }{M }\,  \big( \phi \, \overline{ \ell_{Li}  }  e_{R\alpha} 
+  \phi' \, \overline{ \ell'_{Ri}  }  e'_{L\alpha}\big)   \, + \, {\rm h.c.}
\ee 
Then, if mirror symmetry is exact, i.e. $\langle \phi' \rangle = \langle \phi \rangle = v_{\rm w}$, 
the ordinary and mirror leptons should have identical mass spectra. 
As for neutrinos, now besides the operator (\ref{Yukawa-nu}) generating the neutrino 
Majorrana masses, we should have its mirror copy generating Majorana masses of mirror 
neutrinos, and also a mixed operator between ordinary $\ell$ and mirror $\ell'$ leptons \cite{ABS}:  
\beqn{Yukawa-nu-pr} 
&& \frac{Y_\nu^{ij}  }{\cal M} \, \big(\phi \phi  \, \ell^T_{Li} C \ell_{Lj}  + 
 \phi' \phi'  \, \ell^{\prime T}_{Ri} C \ell'_{Rj}\big) \nonumber \\
&& \quad \, + \, \frac{\tilde{Y}_\nu^{ij}  }{\cal M} \, \phi \phi'  \, \ov{\ell_{Li} } \ell'_{Rj}   
 \, + \, {\rm h.c.}  \, .
\eeqn
The last operator mixes ordinary (active) and mirror (sterile) neutrinos, 
and also can play a key role in co-leptogenesis scenario which can generate 
baryon asymmetries in both ordinary and mirror sectors \cite{BB-PRL}.  
Interestingly, if lepton numbers  (or better $\rB-\rL$ and $\rB'-\rL'$) are conserved 
in each sector, then  these operators are forbidden and all neutrinos remain massless. 
However, if the combination $(\rB\!-\!\rL)+ (\rB'\!-\!\rL')$ is conserved, then the last 
operator is allowed.  In this case the neutrinos will be Dirac particles 
having masses $\sim v_{\rm w}^2/{\cal M}$, with their LH components living  in ordinary world 
and the RH components living in mirror world. 

However, introducing of the mirror fermions does not fully solves the anomaly problem:   
there remains a mixed triangle anomaly of hypercharge--flavor $U(1)\times SU(3)_e^2$. 
For its cancellation, new fermion species should be introduced in the proper representations 
of the  SM  and $SU(3)_e$.  There are several ways of doing this. Let us consider 
one of possibilities  by introducing in our sector, in addition to the regular leptons (\ref{reps}), 
the new lepton species in representations
\be{E-like}
{\cal E}_{L\alpha} \sim (1,-2,3_e;X),    \quad {\cal E}_{Ri} \sim (1,-2,1;X),  
\ee
and, for mirror parity, analogous species in mirror sector: 
\be{Epr-like}
{\cal E}'_{R\alpha} \sim (1,-2',3_e;X),    \quad {\cal E}'_{Li} \sim (1,-2',1;X),  
\ee
where $\alpha=1,2,3$ is a gauge $SU(3)_e$ index and $i=1,2,3$ is just for numbering three 
species.  
We assign to these  fermions a new charge $X$  of additional gauge symmetry 
$U(1)_X$ while ordinary leptons have no $X$-charges. 
  This additional charge  is introduced 
in order to forbid the mixing of new fermions (\ref{E-like}) 
with ordinary leptons (\ref{reps}) due to the mass term 
$M \ov{E_{L\alpha} } e_{R\alpha} $ and the Yukawa terms $\ov{\ell_{Li}} E_{Rj} \phi$ which 
would ruin the flavor structure induced by the operator (\ref{op-leptons}).  
It is easy to check that  by introducing extra fermions (\ref{E-like}) and (\ref{Epr-like}) 
the mixed triangle anomalies including $U(1)\times SU(3)_e^2$, $U(1)_X\times SU(3)_e^2$, 
$U(1)\times U(1)_X^2$ and $U(1)_X\times U(1)^2$ are all cancelled.

The new fermions get masses from couplings with flavons $\xi_n$: 
\be{new-masses} 
y_{in}\xi_n^\alpha   \ov { {\cal E}_{Ri} }{\cal E}_{L\alpha} + 
y_{in}\xi_n^\alpha   \ov { {\cal E}'_{Li} }{\cal E}'_{R\alpha}    + {\rm h.c.}
\ee
where $y_{in}$ are order 1 Yukawa constants. 
Therefore, their mass spectrum should reflect the hierarchy $v_3 : v_2 :v_1 \sim 1 : \eps : \eps\teps$. 
In particular, if $v_2$  is the TeV range, then $v_1$ should be in the range of 100 GeV and 
thus the lightest of new leptons will have a mass of this order. 
In addition, if $U(1)_X$ symmetry is unbroken, then the lightest of these states should be stable 
(the heavier ones will decay in to lighter one via $SU(3)_e$ flavor boson mediated operators). 
Interestingly, the LEP direct experimental lower limit on the mass of new charged leptons is $102.6$ GeV 
\cite{Achard}.  Such heavy leptons can be within the reach of new $e^+ e^-$ machines as ILC/CLIC 
or CEPC/FCC-ee. If $U(1)_X$ symmetry is spontaneously broken, then the mixing of $e,\mu\tau$ 
with new leptons can be allowed and thus the latter can be rendered unstable.

Let us turn to flavor gauge bosons of $SU(3)_3$ which now interact with both normal 
leptons mirror leptons. Now their exchange should create mixed effective operators 
involving both ordinary and mirror leptons. In particular, the bosons $\cF_{1,2,3}^\mu$ 
mediate the following operators involving only first two families of ...
%
\be{G2-mix}
 \frac{1}{v_2^2}   \sum_{a=1}^3 J^\mu_a J'_{a\mu}  = 
\frac{1}{4v_2^2} \sum_{a=1}^3 \, (\ov{\mathbf{e}_R} \, \lambda_a \gamma^\mu \mathbf{e}_R) 
    (\ov{\mathbf{e}'_L} \, \lambda_a \gamma_\mu \mathbf{e}'_L) 
\ee 
Thus, this operator induces muonium - mirror muonium conversion $M(\bar\mu e) \to M'(\mu'\bar e')$ 
with $G_{MM'}/G_F =  (v_{\rm w}/v_2)^2 = 7.6 \times 10^{-3} (2\, {\rm TeV}/v_2)^2$. 
In difference from the muonium-antimuonium conversion (\ref{Muonium}), here is no 
suppression by small mixing angles. 
The present limit on the muonium disappearance reads ${\rm Br}(M \to {\rm invisible}) < 5.7 \times 10^{-6}$ 
\cite{Gninenko} which is respected for $v_2 \sim 400$ GeV or so,   however this limit can be 
improved by several orders of magnitude 
as discussed in Ref. \cite{Gninenko}. 
Analogously, this operator should induce positronium conversion into mirror positronium \cite{Glashow}, 
but for $v_2 > 2$~TeV the positronium disappearance rate is much below the present 
experimental limit Br(o-Ps$\to$invisible) $< 6 \times 10^{-4}$ \cite{Vigo}.

\medskip

\noindent{\bf 8.} 
In this paper we discussed phenomenological implications of horizontal gauge symmetry 
$SU(3)_e$ acting only in lepton sector, between three families of right-handed leptons. The lepton mass hierarchy 
$m_\tau \gg m_\mu \gg m_e$ can be related to the hierarchy of the symmetry breaking 
scales $v_3 \gg v_2 \gg v_1$. We have shown that the LFV effects induced by flavor changing 
gauge bosons are strongly suppressed due to custodial properties of $SU(2)_e\subset SU(3)_e$ 
symmetry and respective scale can be as small $v_2=2$~TeV, which limit is in fact  set 
from the compositeness limits on the flavor-conserving operators while the limits obtained from 
the LFV processes itself are weaker. Taken into account that the gauge coupling constant  
$g$ of horizontal $SU(3)_e$ can be less than 1, then masses of the $SU(2)_e$ 
 gauge bosons $M_{1,2,3} \simeq (g/\sqrt2) v_2$ can be as small as 1 TeV or even smaller, 
and thus can be accessible at new electron-positron machines. 

Analogously to $SU(3)_e$, all $SU(3)$ factors in (\ref{max-1}) can be rendered anomaly 
along the lines discussed in previous section, 
and thus they also can be gauge symmetries.\footnote{
Moreover, also some anomaly free combinations of $U(1)$ factors in (\ref{max}) can be promoted 
 as gauge symmetries, as e.g. common $U(1)_{\rB - \rL}$  acting  between ordinary and mirror sectors 
\cite{Addazi}.  
} 
The quark mass hierarchy can be related with hierarchies in breaking of 
$SU(3)_Q\times SU(3)_d \times SU(3)_u$ gauge factors, 
i.e. with the ratios   $\eps_d=v_2^d/v_3^d$ and $\teps_d=v_1^d/v_2^d$ 
for between the VEVs of $SU(3)_d$ triplet flavons, 
and the same for $SU(3)_u$ and $SU(3)_Q$. In this way, the hierarchy of down quark masses 
will go parametrically as $1 : \eps_d\eps_Q :  \eps_d\teps_d \eps_Q \teps_Q$ and of the up quarks 
as $1 : \eps_u\eps_Q :  \eps_u\teps_u \eps_Q \teps_Q$. 
The quark flavor violating proceses mediated by gauge bosons of $SU(3)_d$ and $SU(3)_u$ 
will be suppressed due to custodial symmetry 
in the same way as the LFV processes mediated by $SU(3)_e$ bosons. 
In particular, the operator $(\ov{s_R}\gamma^\mu e_R)^2$ which induces 
$K^0 -\bar{K}^0$ oscillation (analogously as leptonic operator (\ref{Muonium}) induces $M-\ov M$ conversion) 
will be suppressed by a factor $\sim \eps_d^2 \teps_d^2 \ll 1$. This can allow to quark flavor changing 
gauge bosons to be in the range of few TeV, in fact limited only by the quark compositeness bounds.   
Interestingly, the flavor bosons of $SU(3)_\ell$ and $SU(3)_Q$ can give also anomalous 
contributions imitating the charged current $\times$ current operators of the SM, and so they will have 
interference with the latter.
E.g. $SU(3)_\ell$ bosons induce operator 
$(\ov{e_L} \gamma_\rho \mu_L) (\ov{\nu_\mu} \gamma^\rho \nu_e)$ which is nothing but 
the Fierz-transformed  SM operator (\ref{SM-mu}) responsible for the muon decay. 
Analogously, $SU(3)_Q$ bosons  should induce e.g. operator 
$(\ov{u_L} \gamma_\rho c_L) (\ov{s_L} \gamma^\rho d_L)$ which also interferes with the 
charged current operators in the SM. So, presence of such operators can have impact 
for the unitarity  tests of the CKM mixing of quarks.  Detailed analysis of these issues will be 
given elsewhere \cite{BB-new}. 

As far as the presence of mirror sector is concerned, mirror matters a viable candidate 
for light dark matter dominantly consisting of mirror helium and hydrogen atoms \cite{BCV}. 
The flavor gauge bosons related to both quarks and mirror quarks  
appear as messengers between two sectors and can give an interesting portal 
for direct detection via mirror matter scattering off normal nuclei in detectors \cite{Cerulli}.  
However, they will also give rise to the mixing between 
the neutral ordinary and mirror mesons.  Namely, the lighter flavor bosons induce mixings as  
$\pi^0-\pi^{0\prime}$, $K^0-K^{0\prime}$, etc. \cite{Mirror}, with implications for the invisible decay channels 
of neutral mesons (for a recent discussion, see also Ref. \cite{Emidio}). 
In the supersymmetric version the respective flavor gauginos, complemented by R-parity breaking,  
can induce the mixing between the ordinary and mirror neutral baryons. 
Interestingly, neutron-mirror neutron oscillation $n-n'$  related to physics at the scale of few TeV  \cite{BB-nn'}. 
can  be rather fast, in fact much faster than the neutron decay itself. 
There are some anomalies in existing experiments on $n-n'$ oscillation search \cite{Nesti} 
(Recent summary of experimental bounds on the $n-n'$ oscillation time can be found in Ref. \cite{ILL}). 
which can be tested  in planned experiments on the neutron disappearance and regeneration \cite{ORNL}.

Oscillation between ordinary and mirror neutral particles are effective if they are degenerate in mass, 
i.e. mirror parity is unbroken and the weak scales $\langle \phi \rangle = v_{\rm w}$ and 
$\langle \phi' \rangle = v'_{\rm w}$ are exactly equal in two sectors, $v'_{\rm w} = v_{\rm w}$. 
However, the cancellation of horizontal anomalies between  two sectors does not require 
that mirror parity is unbroken, and in fact one can consider models where it is 
spontaneously broken, e.g. $v'_{\rm w} > v_{\rm w}$, 
with interesting implications for mirror dark matter properties 
and sterile mirror neutrinos \cite{BDM} and axion physics \cite{Giannotti}. In particular, 
 the mirror twin Higgs mechanism for  solving the little hierarchy problem,   
in  supersymmetric \cite{Alice} or non-supersymmetric \cite{Chacko} versions, 
needs $v'_{\rm w}$ in the TeV range, and after all, a scale of few TeV is of interest 
as a realistic scale of consistent supersymmetric models \cite{Miele}. 

In this paper we demonstrated that  in the TeV range there may exist other sorts of new physics  
related to the fermion flavor which can be revealed in future experiments at the energy and precision frontiers.  
In particular, the lepton-flavor changing gauge bosons can be as light as a TeV, or 
even lighter, since the LFV processes are strongly suppressed by custodial symmetry reasons.  
Nevertheless, some of these LFV processes, as e.g. $\tau\to 3mu$,  can have widths close to present 
experimental limits  and can be within the reach of future high precision experiments.   
 

\bigskip 

The preliminary version of this work was presented by B.B. at 
the European Physical Society Conference on High Energy Physics EPS-HEP 2017 \cite{Belfatto}.


\begin{thebibliography}{99}


\bibitem{FCNC} 
  S.~L.~Glashow and S.~Weinberg,
  Phys.\ Rev.\ D {\bf 15}, 1958 (1977); 
  E.~A.~Paschos,
  Phys.\ Rev.\ D {\bf 15}, 1966 (1977).

\bibitem{Froggatt} 
  C.~D.~Froggatt and H.~B.~Nielsen,
  Nucl.\ Phys.\ B {\bf 147}, 277 (1979).

\bibitem{Chkareuli} 
 Z.~Berezhiani and J.~L.~Chkareuli,
  Sov.\ J.\ Nucl.\ Phys.\  {\bf 37}, 618 (1983);   
  JETP Lett.\  {\bf 35}, 612 (1982); 
  JETP Lett.\  {\bf 37}, 338 (1983); 
  Sov.\ Phys.\ Usp.\  {\bf 28}, 104 (1985). 
 
\bibitem{PLB} 
  Z.~Berezhiani,
  Phys.\ Lett.\  B {\bf 129}, 99 (1983); 
  Phys.\ Lett.\  B {\bf 150}, 177 (1985).
  
\bibitem{Khlopov} 
  Z.~Berezhiani and M.~Y.~Khlopov,
  Z.\ Phys.\ C {\bf 49}, 73 (1991); 
  Sov.\ J.\ Nucl.\ Phys.\  {\bf 51}, 739 (1990); 
  Sov.\ J.\ Nucl.\ Phys.\  {\bf 51}, 935 (1990); 
  Sov.\ J.\ Nucl.\ Phys.\  {\bf 52}, 60 (1990); 
  Sov.\ J.\ Nucl.\ Phys.\  {\bf 52}, 344 (1990).  
 
 
\bibitem{MFV1} 
  Z.~Berezhiani,
  Phys.\ Lett.\ B {\bf 417}, 287 (1998) 
[hep-ph/9609342]; 
  Nucl.\ Phys.\ Proc.\ Suppl.\  {\bf 52A}, 153 (1997) 
[hep-ph/9607363].

 
\bibitem{MFV2} 
  A.~Anselm and Z.~Berezhiani,
  Nucl.\ Phys.\ B {\bf 484}, 97 (1997)
 [hep-ph/9605400]; 
  Z.~Berezhiani and A.~Rossi,
  Nucl.\ Phys.\ Proc.\ Suppl.\  {\bf 101}, 410 (2001)
[hep-ph/0107054]; 
  G.~D'Ambrosio, G.~F.~Giudice, G.~Isidori and A.~Strumia,
  Nucl.\ Phys.\ B {\bf 645}, 155 (2002)
 [hep-ph/0207036].
  
 \bibitem{SO10} 
  Z.~Berezhiani and A.~Rossi,
  Nucl.\ Phys.\ B {\bf 594}, 113 (2001) 
 [hep-ph/0003084]; 
  JHEP {\bf 9903}, 002 (1999)
[hep-ph/9811447]; 
  S.~F.~King and G.~G.~Ross,
  Phys.\ Lett.\ B {\bf 574}, 239 (2003)
  [hep-ph/0307190]; 
  Z.~Berezhiani and F.~Nesti,
  JHEP {\bf 0603}, 041 (2006)  
[hep-ph/0510011].

\bibitem{U1A}
  L.~E.~Ibanez and G.~G.~Ross,
  Phys.\ Lett.\ B {\bf 332}, 100 (1994)
  [hep-ph/9403338]; 
  V.~Jain and R.~Shrock,
  Phys.\ Lett.\ B {\bf 352}, 83 (1995)
  [hep-ph/9412367];  
  P.~Binetruy, S.~Lavignac and P.~Ramond,
  Nucl.\ Phys.\ B {\bf 477}, 353 (1996)
  [hep-ph/9601243]; 
  E.~Dudas, C.~Grojean, S.~Pokorski and C.~A.~Savoy,
  Nucl.\ Phys.\ B {\bf 481}, 85 (1996)
  [hep-ph/9606383]; 
  Z.~Berezhiani and Z.~Tavartkiladze,
  Phys.\ Lett.\ B {\bf 396}, 150 (1997)
  [hep-ph/9611277]; 
  Phys.\ Lett.\ B {\bf 409}, 220 (1997)
  [hep-ph/9612232].


\bibitem{BB-new} 
B. Belfatto and Z. Berezhiani, in preparation. 

\bibitem{Dvali} 
  Z.~Berezhiani and G.~Dvali,
  Phys.\ Lett.\ B {\bf 450}, 24 (1999) 
  [hep-ph/9811378].

\bibitem{Cahn} 
  R.~N.~Cahn and H.~Harari,
  Nucl.\ Phys.\ B {\bf 176}, 135 (1980).
 
\bibitem{Low} 
  Z.~G.~Berezhiani and J.~L.~Chkareuli,
  Sov.\ J.\ Nucl.\ Phys.\  {\bf 52}, 383 (1990). 

\bibitem{Weinberg} 
  S.~Weinberg,
  Phys.\ Rev.\ Lett.\  {\bf 43}, 1566 (1979).

\bibitem{Bourilkov:2001pe} 
  D.~Bourilkov,
  Phys.\ Rev.\ D {\bf 64}, 071701 (2001)  
[hep-ph/0104165].
  
\bibitem{Schael:2006wu} 
  S.~Schael {\it et al.} [ALEPH Collaboration],
  Eur.\ Phys.\ J.\ C {\bf 49}, 411 (2007)
  [hep-ex/0609051].

\bibitem{PDG} 
  M.~Tanabashi {\it et al.} [Particle Data Group],
  Phys.\ Rev.\ D {\bf 98}, no. 3, 030001 (2018).

\bibitem{Feinberg} 
  G.~Feinberg and S.~Weinberg,
  Phys.\ Rev.\  {\bf 123}, 1439 (1961).

\bibitem{Willmann} 
  L.~Willmann {\it et al.},
  Phys.\ Rev.\ Lett.\  {\bf 82}, 49 (1999) 
 [hep-ex/9807011].
 

\bibitem{ae}
  P.~Foldenauer and J.~Jaeckel,
  JHEP {\bf 1705}, 010 (2017). 

 \bibitem{Parker}
R.~H.~Parker {\it et al.,} 
  Science {\bf 360}, 191 (2018). 

\bibitem{Mirror} 
  T. D. Lee and C. N. Yang, 
 Phys.\ Rev.  {\bf 104}, 254 (1956);  
  I.~Y.~Kobzarev, L.~B.~Okun and I.~Y.~Pomeranchuk,
  Yad.\ Fiz.\  {\bf 3}, 1154 (1966); 
  R.~Foot, H.~Lew and R.~R.~Volkas,
 Phys.\ Lett. \ B {\bf 272} (1991) 67. 

\bibitem{Alice} 
 Z.~Berezhiani,
  ``Through the looking-glass: Alice's adventures in mirror world,''
 In ``From Fields to Strings: Circumnavigating Theoretical Physics", Eds.  M.  Shifman {\it et al.}: 
 Vol. 3, pp.  2147-2195  
  [hep-ph/0508233].  


\bibitem{Alice2} 
  Z.~Berezhiani,
  Int.\ J.\ Mod.\ Phys.\ A {\bf 19}, 3775 (2004)
[hep-ph/0312335]; 
  Eur.\ Phys.\ J.\ ST {\bf 163}, 271 (2008).
For a historical overview, see  
L.~B.~Okun,
 Phys.\ Usp.\  {\bf 50}, 380 (2007)
 [hep-ph/0606202].


\bibitem{ABS} 
  E.~K.~Akhmedov, Z.~G.~Berezhiani and G.~Senjanovic,
  Phys.\ Rev.\ Lett.\  {\bf 69}, 3013 (1992); 
  R.~Foot, H.~Lew and R.~R.~Volkas,
  Mod.\ Phys.\ Lett.\ A {\bf 7}, 2567 (1992); 
  R.~Foot and R.~R.~Volkas,
  Phys.\ Rev.\ D {\bf 52}, 6595 (1995)
  [hep-ph/9505359]; 
Z.~Berezhiani and R.~N.~Mohapatra,
  Phys.\ Rev.\ D {\bf 52}, 6607 (1995) 
[hep-ph/9505385].

\bibitem{BB-PRL}
  L.~Bento and Z.~Berezhiani,
  Phys.\ Rev.\ Lett.\  {\bf 87}, 231304 (2001)
  [hep-ph/0107281]; 
  Fortsch.\ Phys.\  {\bf 50}, 489 (2002); 
  hep-ph/0111116.

\bibitem{Achard} 
  P.~Achard {\it et al.} [L3 Collaboration],
  Phys.\ Lett.\ B {\bf 517}, 75 (2001)
  [hep-ex/0107015].

\bibitem{Addazi}
  Z.~Berezhiani,
Eur.\ Phys.\ J.\ C {\bf 76}, no. 12, 705 (2016)
  [arXiv:1507.05478 [hep-ph]]; 
  K.~S.~Babu and R.~N.~Mohapatra,
  Phys.\ Rev.\ D {\bf 94}, no. 5, 054034 (2016)
  [arXiv:1606.08374 [hep-ph]]; 
  A.~Addazi, Z.~Berezhiani and Y.~Kamyshkov,
  Eur.\ Phys.\ J.\ C {\bf 77}, no. 5, 301 (2017)
  [arXiv:1607.00348 [hep-ph]]. 

\bibitem{Gninenko} 
  S.~N.~Gninenko, N.~V.~Krasnikov and V.~A.~Matveev,
  Phys.\ Rev.\ D {\bf 87}, no. 1,  015016 (2013).

\bibitem{Glashow} 
S.~L.~Glashow,
  Phys.\ Lett.\  {\bf 167B}, 35 (1986); 
S.~N.~Gninenko,
  Phys.\ Lett.\ B {\bf 326}, 317 (1994).

\bibitem{Vigo} 
  C.~Vigo, L.~Gerchow, L.~Liszkay, A.~Rubbia and P.~Crivelli,
  Phys.\ Rev.\ D {\bf 97}, no. 9, 092008 (2018)
  [arXiv:1803.05744 [hep-ex]].

\bibitem{BCV}
  Z.~Berezhiani, D.~Comelli and F.~L.~Villante,
  Phys.\ Lett.\ B {\bf 503}, 362 (2001)
  [hep-ph/0008105]; 
  A.~Y.~Ignatiev and R.~R.~Volkas,
  Phys.\ Rev.\ D {\bf 68}, 023518 (2003)
  [hep-ph/0304260]; 
  Z.~Berezhiani, P.~Ciarcelluti, D.~Comelli and F.~L.~Villante,
  Int.\ J.\ Mod.\ Phys.\ D {\bf 14}, 107 (2005)
  [astro-ph/0312605]; 
  Z.~Berezhiani, S.~Cassisi, P.~Ciarcelluti and A.~Pietrinferni,
  Astropart.\ Phys.\  {\bf 24}, 495 (2006)
  [astro-ph/0507153].

 \bibitem{DAMA} 
  R.~Cerulli {\it et al.}, 
  Eur.\ Phys.\ J.\ C {\bf 77}, no. 2, 83 (2017)
  [arXiv:1701.08590 [hep-ex]]; 
A.~Addazi {\it et al.},  
  Eur.\ Phys.\ J.\ C {\bf 75}, no. 8, 400 (2015)
  [arXiv:1507.04317 [hep-ex]].
 


\bibitem{Emidio} 
  D.~Barducci, M.~Fabbrichesi and E.~Gabrielli,
  Phys.\ Rev.\ D {\bf 98}, no. 3, 035049 (2018)
  [arXiv:1806.05678 [hep-ph]].
  
\bibitem{BB-nn'} 
  Z.~Berezhiani and L.~Bento,
  Phys.\ Rev.\ Lett.\  {\bf 96}, 081801 (2006)
  [hep-ph/0507031]; 
  Phys.\ Lett.\ B {\bf 635}, 253 (2006)
  [hep-ph/0602227]; 
  Z.~Berezhiani,
  Eur.\ Phys.\ J.\ C {\bf 64}, 421 (2009)
  [arXiv:0804.2088 [hep-ph]]. 
  
  
\bibitem{Nesti}
  Z.~Berezhiani and F.~Nesti,
  Eur.\ Phys.\ J.\ C {\bf 72}, 1974 (2012)
  [arXiv:1203.1035 [hep-ph]].
  
\bibitem{ILL} 
  Z.~Berezhiani {\it et al.}, 
  Eur.\ Phys.\ J.\ C {\bf 78}, no. 9, 717 (2018)
  [arXiv:1712.05761 [hep-ex]].


  \bibitem{ORNL}
  Z.~Berezhiani, M.~Frost, Y.~Kamyshkov, B.~Rybolt and L.~Varriano,
  Phys.\ Rev.\ D {\bf 96}, 
  035039 (2017)
  [arXiv:1703.06735 [hep-ex]]; 
  L.~J.~Broussard {\it et al.},
  arXiv:1710.00767 [hep-ex].


\bibitem{BDM} 
  Z.~Berezhiani, A.~D.~Dolgov and R.~N.~Mohapatra,
  Phys.\ Lett.\ B {\bf 375}, 26 (1996) 
[hep-ph/9511221];
  Z.~Berezhiani,
  Acta Phys.\ Polon.\ B {\bf 27}, 1503 (1996) 
 [hep-ph/9602326].  
 
\bibitem{Giannotti} 
  Z.~Berezhiani, L.~Gianfagna and M.~Giannotti,
  Phys.\ Lett.\ B {\bf 500}, 286 (2001)
  [hep-ph/0009290].
 
\bibitem{Chacko} 
  Z.~Chacko, H.~S.~Goh and R.~Harnik,
  Phys.\ Rev.\ Lett.\  {\bf 96}, 231802 (2006)
  [hep-ph/0506256].
 
\bibitem{Miele} 
  Z.~Berezhiani, M.~Chianese, G.~Miele and S.~Morisi,
  JHEP {\bf 1508}, 083 (2015)
  [arXiv:1505.04950 [hep-ph]].
 
  
\bibitem{Belfatto} 
  B.~Belfatto,
  PoS EPS {\bf -HEP2017}, 660 (2017).

\end{thebibliography}
\end{document}